\input{aipcheck}

\documentclass[
,final            % use final for the camera ready runs
  ,numberedheadings % uncomment this option for numbered sections
  ]
  {aipproc}

\layoutstyle{6x9}

\begin{document}

\title{Type Ia Supernova Rates Near and Far}

% <Replace this text with PACS numbers; choose from this list:
%                \texttt{http://www.aip..org/pacs/index.html}>}

\author{Nino Panagia}{ address={STScI, 3700 San Martin Dr., Baltimore, MD 21218,
USA; \texttt{panagia@stsci.edu}} , altaddress={Istituto Nazionale di Astrofisica
(INAF), Via del Parco Mellini 84, I-00136, Rome, Italy; and \\Supernova
Ltd., OYV \#131, Northsound Road, Virgin Gorda, British Virgin
Islands.}
}

\author{Massimo Della Valle}{
  address={INAF - Osservatorio di Arcetri, Largo E. Fermi 5, I-50125
  Firenze, Italy ; \texttt{massimo@arcetri.inaf.it}  }
}
\author{Filippo Mannucci}{
  address={INAF - Istituto di Radioastronomia, Largo E. Fermi 5, I-50125
  Firenze, Italy; \texttt{filippo@arcetri.inaf.it}  }
}

\begin{abstract}

Recently, three important observational results were established: {\it
(a)} The evolution of the SNIa rate with redshift is now measured up to
z$\sim$1.6 and  the results at the highest redshifts, derived by the
GOODS collaboration show that the SN rate rises up to z$\sim$0.8, when
the Universe was 6.5 Gyr old, and decreases afterward. ~~{\it (b)} The
rate of supernova explosions of the different types as a function of the
galaxy (B-K) and the galaxy mass have been determined. It is found that
the rates of all  SN types, including Ia, Ib/c and II, show a marked
increase with the star formation activity. ~~{\it (c)} An analysis of
SNIa events in early-type galaxies has provided conclusive evidence that
the rate of SNIa in radio-loud galaxies is much higher than the rate
measured in radio-quiet galaxies. This result suggests that repeated
episodes of interaction and/or mergers of early-type galaxies with dwarf
companions are responsible for supplying an adequate number of SNIa
progenitors to the stellar population of elliptical galaxies. 

On this basis we have discussed the distribution of the delay time (DTD)
between the formation of a SNIa progenitor star and its explosion as a
SNIa. Our analysis finds: i) models with long delay times, say  3-4 Gyr,
cannot reproduce the dependence of the SNIa rate on the colors and on
the radio-luminosity of the parent galaxies; ii) the dependence of the 
SNIa rate on the parent galaxy colors requires models with a wide DTD,
spanning the interval 100 Myr to 10 Gyr; iii) the dependence on the
parent galaxy radio-luminosity requires substantial production of SNIa
at epochs earlier than 100 Myr after the birth of a given stellar
generation; iv) the comparison between observed SN rates and a grid of
theoretical "single-population" DTDs shows that only a few of them are
marginally consistent with all observations; v) the present data are
best matched by a bimodal DTD, in which about 50\% of type Ia SNe ({\it
"prompt"} SNIa) explode soon after their stellar birth, in a time of the
order of 100 Myrs, while the remaining 50\% ({\it "tardy"} SNIa) have a
much wider distribution, well described by an exponential function with
a decay time of about 3 Gyr. This fact, coupled with the well
established bimodal distribution of the decay rate, suggests the
existence of two classes of progenitors and/or explosive channels. We
discuss the cosmological implications of this result and make simple
predictions.

\end{abstract}

\maketitle

%%%%%%%%%%%%%%%%%%%%%%%%%%%%%%%%%%%%%%%%%%%%
%% MAINMATTER
%%%%%%%%%%%%%%%%%%%%%%%%%%%%%%%%%%%%%%%%%%%%
\section{Introduction}
Type Ia supernovae (SNe) are very important objects in modern cosmology
because they are bright sources that can be detected up to large
distances and it appears that their intrinsic luminosities can be
inferred directly from  their light curves. Exploiting these properties,
the study of SNIa at high redshifts has allowed the discovery of the
cosmic acceleration (Perlmutter et al. 1998, Riess et al. 1998,
Perlmutter et al. 1999).  Even though these objects are commonly
believed to be associated with the explosion of a degenerate star as a
white dwarf (e.g. Hillebrandt \& Niemeyer 2000), the nature of SNIa
progenitors is not firmly established, and several explosion patterns
are possible  (see, e.g., Branch et al. 1995, Yungelson 2004) and each
of these may dominate at different redshifts.  As a consequence, the
existence of systematics affecting SNIa at different redshifts cannot be
ruled out  (e.g. Kobayashi et al., 1998; Nomoto et al., 2003) and it is
worth being further investigated for possible cosmological implications.

Different explosion models (e.g. Greggio \& Renzini 1983, Yungelson \&
Livio 2000, Matteucci \& Recchi 2001,  Belczynski, Bulik \& Ruiter 2005,
Greggio 2005) predict different delay times between the formation of the
progenitor system and the SN explosion.  Differences in the expected
delay times are  testable with the observations (e.g. Madau, Della Valle
\& Panagia 1998, Sadat et al. 1998, Dahlen \& Fransson 1999) so that
constraining the Delay Time Distribution (DTD) will permit one to
ascertain the nature of SNIa progenitors by confirming or excluding some
of these models.
%
%-------------------------------------------------------------------------
\section{New observational evidence}
 
In the last few years, three important observational results were established:

(1) The evolution of the SNIa rate with redshift is now measured up to
z$\sim$1.6 (Hardin et al. 2000, Pain et al. 2002, Strolger  2003,
Madgwick et al. 2003, Cappellaro et al. 2004, Gal-Yam \& Maoz 2004,
Mannucci et al. 2005, Barris \& Tonry 2006, Neill et al. 2006).  The
results at the highest redshifts, derived by the GOODS collaboration
(Dahlen et al. 2004, Strolger et al. 2004, 2005)  show that the SN rate
rises up to z$\sim$0.8, when the Universe was 6.5 Gyr old (see panel b
of Fig.~1), and decreases afterward.  This behavior can be compared with
the cosmic Star Formation History (SFH) which continues to rise up to
z$\sim2.5$, i.e., at a time about 4 Gyr earlier  (Madau, Pozzetti \&
Dickinson 1998, Giavalisco et al. 2004). These results have been
interpreted  by Dahlen et al. (2004) and Strolger et al (2005) as
evidence of a very long delay time ($\sim3-4$ Gyr)  between the
formation of the stars in the binary system and  the explosion of a
SNIa.

(2) Recently, we have determined the SN rates per unit mass in the local
Universe (Mannucci et al. 2005), finding a very strong dependence of the
rates on the (B$-$K) colour of the parent galaxies:  blue galaxies (the
latest Hubble types) exhibit  a SNIa rate  a factor of $\sim 30$  higher
than that of red galaxies (early types). This result indicates that the
delay time must have a wide  distribution.   In star forming galaxies,
the delay time must be at least as short as the timescale of colour
evolution ($\sim 0.5$ Gyr),  while the existence of supernovae in 
galaxies without any recent star formation argues that some SNIa have
long delay times. For this reason Mannucci et al. (2005),  following
earlier suggestions (e.g., Dallaporta 1973, Della Valle \& Livio 1994,
Panagia 2000), have proposed the existence of two populations of
progenitors, one related  to the young stellar  population, with rates
proportional to the recent star formation rate (SFR), the other related
to the old populations, with rates proportional to the total stellar 
mass accumulated over time, i.e. the integral of the SFR over time.

(3) Della Valle et al. (2005) demonstrated that early-type  radio-loud
galaxies show a strong enhancement, by a factor of about 4, of the SNIa
rate with respect to the radio-quiet sample, and that the hypothesis of 
an equal  rate between the samples can be  rejected at a 99.96\%
confidence level (see also Della Valle \& Panagia 2003).  Both the radio
activity and the SN rate enhancement are interpreted in terms of
episodes of star formation due to merging with small galaxies.  Since
each episode of radio activity is estimated to lasts about $10^8$ years
(Srianand \& Gopal-Krishna 1998, Wan, Daly \& Guerra 2000),  the
evolutionary time for most SNIa in radio-loud galaxies must also be
around 100 million years.

All the above issues constitute observational links between the epochs
of star formation and SN explosion, and, therefore, can be used to 
constrain the DTD over different timescales in that: (i)  The evolution
of the SNIa rate with cosmic time is sensitive to long timescales (up to
several Gyr). (ii) The  dependence of the local rate with the parent
galaxy colour samples timescales of the order of the colour evolution of
the galaxies, i.e., up to 0.5-1 Gyr. (iii) The relation between SN rate
and radio power gives information on the timescales of the order of
$10^8$ years, corresponding to the radio activity lifetime.

\section {Single Population DTD}

Looking for a DTD that satisfies all constraints, we have investigated a
large number of possible  ``single-population'' models, i.e., DTDs that
can be associated to a single progenitor population and be described by
a single analytical law (Mannucci, Della  Valle \& Panagia, 2006). We
used DTDs characterized by different shapes (exponential decline,
gaussian shape, and constant over one Hubble time) and characteristic
times between 0.1 and 6 Gyr. None of these simple DTDs can satisfy all
of the observational constraints  simultaneously. Within this class of
models, the observations are best matched by an exponential distribution
of delay times with e-folding time of 3 Gyr. This distribution provides
a rather good description of the observed rates as function of redshift
and of the parent galaxy colours. However, this model is unable to
describe satisfactorily the variation of the rates with the radio-power
of the parent galaxy. These results indicate that while a DTD that
extends over several Gyr is needed, an additional  contribution at early
times (below $10^8$ yr),  is necessary to explain all observations.

Some of the single-degenerate and double-degenerate models, which
predict very broad DTDs provide interesting results. This is the case
for a number of Greggio (2005) models, both single-degenerate (SD) and
double-degenerate (DD),  Yungelson \& Livio (2000) DD Chandrasekar mass
model, and the Matteucci \& Recchi (2001) SD model.  In all these models
the DTD peaks at about $0.6-2\times10^8$ yr and then decays rapidly, 
roughly like $t^{-1}$, dropping by a factor of 10 after about 1 Gyr. 
The dependence of the rates with redshift and galaxy colours are
satisfactorily reproduced, although in some cases the fast evolution
tends to under-predict  the SNIa rate in the reddest galaxies. However,
these DTDs predict about 5-15\% of SNIa to explode within the first
$10^8$ yrs. As a consequence, they produce a SNIa rate in radio loud
galaxies only 10-40\% higher then in radio-quiet galaxies, instead of
the observed factor of 4.  Even if these models cannot be ruled out with
an high degree of confidence, it is clear that a DTD with both more SNIa
explosions at early times and a slower  evolution afterward is needed to
fully account for the observations.

%----------------------------------------------------------------
% figure1
\begin{figure}
\centerline{\includegraphics[width=7.3cm]{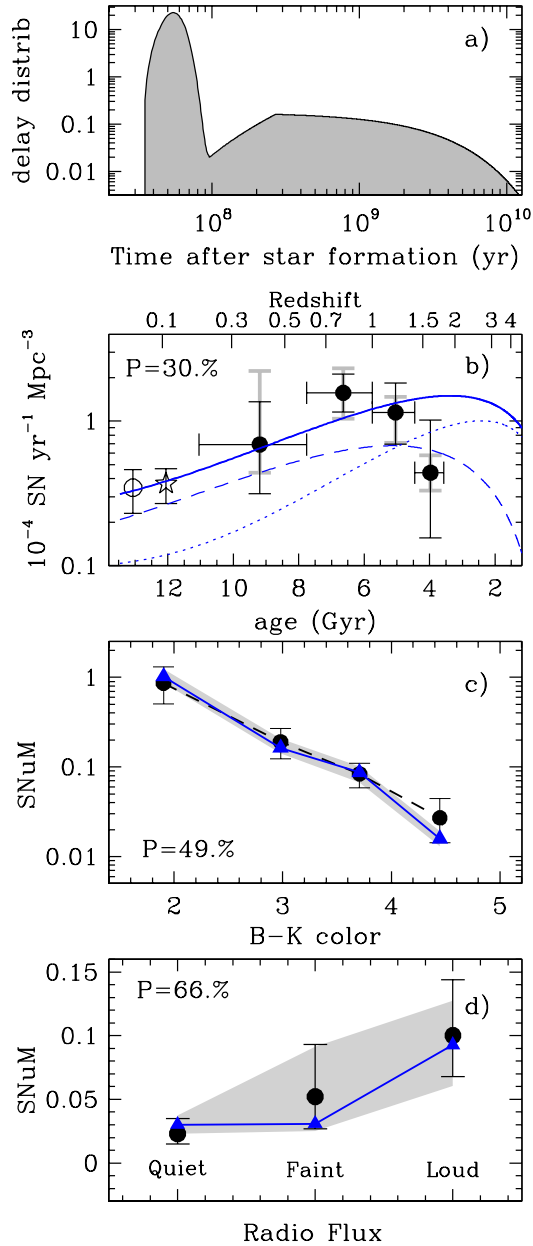} }
\caption{
The SN rates for a DTD constituted by equal
contributions (50\%) of an exponentially declining function with
e-folding time of 3 Gyr  and a gaussian centered at $5\times10^7$ yr 
and $\sigma=10^7$ yr. {\em Panel (a): } the DTD itself  (number of SNIa
per unit time after star formation);  {\em Panel (b):} the evolution of
the rate along the cosmic age.   The solid curve is the prediction  for
the considered DTD.   Data are from Mannucci et al. (2005; open circle),
Strolger et al. (2004; star),  Dahlen (2004; dots,  with black error
bars for1$\sigma$ statistical errors, and the gray bars for systematic
uncertainties).   In panel (b) the dotted and dashed lines show the
contributions from the ``prompt'' and ``tardy'' components,
respectively.  P is the statistical probability of agreement estimated
from the statistical errors only. {\em Panel (c):} the predictions
(solid line and triangles) for SNIa rates as a function of  the parent
galaxy (B$-$K) colour,  expressed in SNe per century per $10^{10}$
M$_\odot$  of stellar mass (SNuM).     The dots and the dashed line show
the observational data from Mannucci et al. (2005).  {\em Panel (d):}
SNIa rates in early-type galaxies as a function of  the radio power of
the parent galaxy. The black dots are the observed SN rates in Della
Valle et al. (2005) with 1$\sigma$ Poisson error.  }
\end{figure}
%----------------------------------------------------------------

\section{Two populations DTDs}
\label{sec:twopop}

For these reasons we considered a set of ``two populations'' models in
which the DTD is obtained as the sum of two distinct functions. In all
cases, we added a ``prompt'' gaussian centered at $5\times10^7$ yr to a
much slower function, either another gaussian or an exponentially
declining function.  We will refer to the former component as ``prompt''
exploders and to latter as ``tardy'' ones\footnote{We chose to  use
``tardy", rather than``delayed", to avoid any misunderstanding with the
``delayed'' detonation model adopted for type Ia supernovae, e.g.
Woosley (1990), or more recently, Golombek \& Niemeyer (2005)}

Figure~1 shows the results of a model in which 50\% of the SNe derive
from the ``prompt'' population, and the remaining 50\% from the
``tardy'' one that consists in an exponentially declining function  with
an e-folding time of 3 Gyr.   These results shows that the observational
data are better reproduced by a DTD with a peak at short times  (below
$10^8$ yr) that includes about half of the SNIa events, and an extension
toward very long times, say, 3 Gyr and beyond. Provided that the
``tardy'' component extends well beyond 3 Gyr, its shape is not well 
constrained by the fit: exponential decays with characteristic times 
between 2.5 and 8 Gyr can still provide reasonable fits. These
uncertainties will reduce considerably when the SNIa rates at z$>$0.5
will be measured with greater accuracy.

We note that a bimodal distribution of delay times should  not be
regarded as just a heuristic method to fit the data,  because there are
theoretical models which actually predict a bimodal DTD.  One of the
best examples (see Figure~7 in Mannucci et al 2006) is provided by 
Belczynski et al. (2005) SD model with reduced common envelope
efficiency ($\alpha\lambda=0.3$).  This model predicts a bimodal DTD
which peaks at $10^8$ and $3\times10^9$ yr, and includes both He and
C-O white dwarf explosions.  This model correctly reproduces the
evolution of the rate with redshift and its dependence on the colours,
but  accounts for the enhancement in the radio-loud galaxies only
qualitatively, because its ``prompt''  peak is centered at $10^8$ yr
instead at the best-fitting value of  $5\times10^7$ yr.  

Bimodal DTDs can also be produced by models with more than one type of
progenitors, for example in which both the single-degenerate and
double-degenerate channels are active (see, for example, Nomoto et al.,
2003).  A bimodal DTD is also naturally produced by the SD model by
Kobayashi et al. (1998) in which two different companion stars are
present: either a red giant with initial mass of about 1 M$_\odot$ \ and
orbital periods of tens to hundreds days,  or a main-sequence star with
mass $\sim2-3$ M$_\odot$ \ and periods  of the order of a day.\\

%-------------------------------------------------------------------------

\section{Discussion}

We have shown that the observational constraints to the SNIa rates,
namely the rate evolution with redshift, the dependence of SNIa rates
with host galaxy colors, and the marked increase of SNIa rates in
radio-loud Ellipticals, are best reproduced if about half of the SNe
explode within $10^8$ yr from star formation (``prompt'' component)
while the rest have explosion timescales of a few Gyr (``tardy''
component). We would like to stress that while the exact shape of these
two distributions cannot be determined accurately, the requirement of
50\% {\it prompt} and 50\% {\it tardy} SNIa is unavoidable. 

Similar conclusions have been reached by Sullivan et al. (2006) from an
analysis of  a sample of 124 SNIa from the Supernova Legacy Survey
(SNLS) distributed over 0.2$<z<$0.75. They also found that passive
galaxies, with no star formation, preferentially host faster
declining/dimmer SNIa, while brighter events are found in systems with
ongoing star formation.

\subsection{Bimodality and stellar mass}
\label{sec:mass}

We note that the main sequence lifetime is about $5\times10^8$ yr for
a star of 3M$_\odot$ , $10^8$ yr for 5.5M$_\odot$ \ and about $4\times10^7$ yr
for 8M$_\odot$ \ (e.g., Girardi et al. 2000). Therefore, the SNe of the
``prompt'' peak, which include about 50\% of the total number of SNIa
events and explode within $10^8$ yr from star formation, must all
derive from stars with masses above 5.5M$_\odot$ . Also, for a Salpeter
IMF, the number of stars between 5.5 and 8M$_\odot$ \ are about a third of
those between 3 and 5.5M$_\odot$ . This implies that the SNIa efficiency
for higher mass progenitors (5.5--8M$_\odot$ ) is about 3 times higher that
for lower mass progenitors (3--5.5M$_\odot$ ). Therefore, given an
overall efficiency of 4.5\%, it follows that the efficiency for higher
mass stars is $\eta$(5.5--8M$_\odot$ )$\sim$6.8\%, and the one for lower
mass stars is $\eta$(3--5.5M$_\odot$ )$\sim$2.3\%.  Therefore, the
requirement that about 50\% of the SNIa explode within $10^8$ yrs
implies that {\em the  efficiency and the characteristic delay time are
expected to change considerably for stellar masses around 5.5M$_\odot$
}.

It is important to realize that the explosion efficiency of the
``prompt'' SNIa is determined unambiguously by their number and the mass
range of the progenitors {\em as directly implied by the observations}.
On the other hand, if one allows the remaining 50\% SNIa, i.e. the
``tardy"  ones, to arise also from stars with masses lower than
3M$_\odot$ , then their inferred explosion efficiency would also
decrease, because the available pool of stars would increase whereas the
number of ``tardy" SNe does not.

We cannot draw conclusions on whether the change of the efficiency at
about $5.5$M$_\odot$ \ is due to a different physical process (e.g., SD
vs. DD) or to one and the same process  operating in separate regions of
the parameter space (e.g., systematic differences of the binary systems
as a function of the stellar mass).  For example,  it could be that the
binary fraction for primary stars with masses above 5.5M$_\odot$ ~ is
markedly higher than for lower mass stars. Or it could be that the
distribution  of secondary star masses is more skewed toward masses
close to the primary star mass and, therefore, the mass transfer be more
efficient and faster (see Pinsonneault \& Stanek 2006 for a discussion).

The currently existing models (e.g. Greggio 2005, Belczynski et al.
2005, Nomoto et al. 2003) are not able to resolve this ambiguity because
of both uncertainties in the model assumptions and possible coexistence
of different physical processes. However, we are confident that a
judicious analysis of data obtained for a large sample of SNIa over a
suitably wide interval of redshifts will make it possible to clarify
this issue.

\subsection{Consequences of the bimodality on Cosmology}

In addition to providing essential clues to the nature of SNIa
progenitors, our results have also important  implications for
cosmology:
\noindent {\bf -} {\em The fractions of SNe coming from the two
populations change with cosmic time}, as can be seen from Figure~1: 
the ``tardy'' SNe dominate at z$<$1.3 and the ``prompt'' SNe  above this
limit. The ratio of the ``prompt'' SN rate to that of the ``tardy'' SNe
changes from 0.5 in the local Universe to about 1.2 at z=1.5. Similar
results are obtained for Belczynski et
al. (2005) SD model.
  
\noindent {\bf -} It is conceivable that the two populations of SNe can
be distinguished also by some intrinsic properties. As an example, it is
possible that ``prompt'' SNIa are, on average, more affected by dust
extinction than the ``tardy'' component, as they must explode closer  to
the formation cloud (e.g. Sullivan et al. 2003, Mannucci, Della Valle \&
Panagia 2007). In this case,  {\em the average properties of SNIa are
expected to change with redshift}, especially at z$>1$ when the
``prompt'' SNe become more common.  The Hubble diagrams used to derive
information on the cosmological parameters (e.g. Riess et al. 2004) are,
up to now, mostly based on SNe at z$<1$ and, therefore, are expected  to
be dominated by the ``tardy'' population. As the ratio between the two
different flavors of type SNIa changes with cosmic time,  {\em
evolutionary effects should become more important at higher redshifts}
(Riess \& Livio, 2006). 

\noindent {\bf -} The luminosity-decline rate relation for SNIa
(Pskovskii 1977, Phillips 1993, Hamuy et al. 1996, Phillips et al. 1997)
is derived in the local Universe and, therefore, under this scenario, is
dominated by the ``tardy'' SNe. The evidence for a cosmological
acceleration relies on the assumption that the same relation holds also
at high redshift (see Rowan-Robinson 2002 and Leibundgut 2004 for a
discussion). If the two populations follow slightly different
relations,  a bias is expected to emerge as a function of redshift,
especially when the ``prompt'' population becomes  dominating, at $z\sim
1.2$.  Thus, a reliable use of SNIa for cosmology measurements at
z$>1$ would require a good understanding of the differences in
properties of the two populations.

\subsection{Bimodality and metallicity evolution}

The existence of two populations of SNIa has direct consequences also
on the chemical evolution of the Universe:

\noindent {\bf -} ``Prompt'' SNIa, having a redshift distribution
similar to the CC SNe, dominate the SNIa population at high redshifts.
Therefore, {\em in the early Universe the production of Fe  is expected
to follow that of Oxygen}, and the O/Fe abundance ratio is expected to
be relatively constant but appreciably higher than in the local
Universe. When  the SNIa ``tardy'' component starts dominating, i.e. 
past the SFH peak at z$\sim2$, the Fe production is boosted and the O/Fe
ratio is expected to decrease rapidly to approach the ``solar" values 
around redshifts $<\sim0.5$. These aspects have recently been discussed
in some detail by Scannapieco \& Bildsten (2005) who  for their model
calculations  adopted the simplified description of the SNIa rates as
derived by Mannucci et al. (2005) in terms of a component proportional
to the star formation rate (SFR) and another one that is proportional to
the  total stellar mass.  More recently, Matteucci et al. (2006) have
included the bimodal DTD for SNIa in chemical evolution model
calculations, to find that  this scenario is fully consistent with the
main chemical properties of galaxies of various morphological types.

\noindent {\bf -}  It is known that the intra-cluster medium is
relatively rich in iron ([Fe/H]$\sim-0.5$) and that the metallicity
shows a very mild evolution with redshift (Tozzi et al. 2003). The
observed iron mass is about a factor of 6 larger than  could have been
produced by core-collapse SNe (Maoz \& Gal-Yam 2004)  and a factor of 10
larger than that produced by the current rate  of SNIa (Renzini, 2004).
The ``two populations'' model naturally explains  these observations, as
the current type Ia rate is just the long-time  declining tail of a SN
distribution that peaked at early cosmic times. The amount of observed
iron and its redshift evolution is reproduced by assuming an average age
of the  stars in clusters of 10 Gyr (see Matteucci et al. 2006).

%----------------------------------------------------------------
% figure2
\begin{figure}
\centerline{\includegraphics[width=7.9cm]{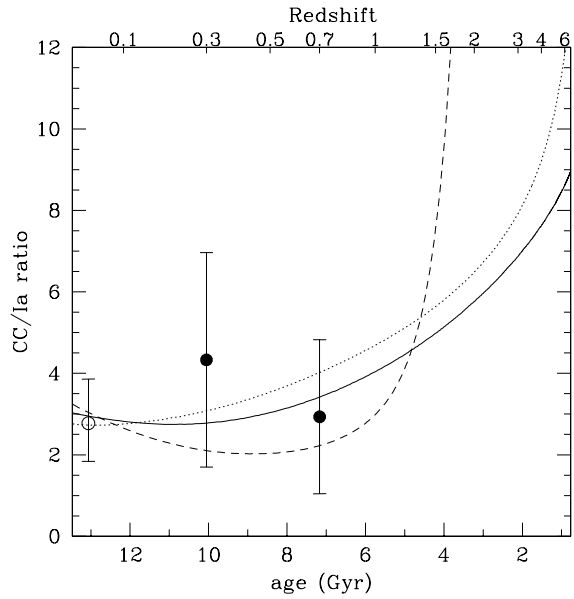} }
\caption{
Ratio of the rates of the CC to Ia SNe as a function of the redshift.
The white and black dots are observed values (Mannucci et al. 2005,
Dahlen et al. 2004).  The lines show the predictions of the gaussian
``single-population'' model (i.e. a model in which the DTD is a narrow
gaussian centered at 3-4 Gyr;  dashed line), Yungelson and Livio
(2000) DD Chandrasekar mass model (dotted line), and the
``two-populations'' shown in Figure~1 (solid line).  The predictions use
a Salpeter IMF  and mass ranges of 3-8M$_\odot$ \ (SNIa) and
8-40M$_\odot$ \ (CC SNe), and are scaled to match the observed values.
}
\end{figure}
%----------------------------------------------------------------

\subsection{Predictions}

The existence of the ``prompt'' and ``tardy'' populations of type Ia SNe
can be tested by two observations:

\noindent $\bullet$ The SNIa rate is expected not to decrease
significantly  moving toward high redshifts up to z$\sim$2,  at which
the cosmic star formation history has its broad peak.  As a consequence,
{\em it should be possible to detect SNIa up to high redshifts, say,
z$\sim5$} so as to discriminate among  different cosmological models. 
In particular, at z$>$2 the SNIa rate should be nearly constant at a
level of about $10^{-4}$~SN~yr$^{-1}$~Mpc$^{-3}$ (see Figure~1).  Such a
rate can be reduced only if the effects of metallicity evolution become
important at z$\sim$1, and if the changing in metallicity has an
important effect in the explosion rate as predicted by Kobayashi et al.
(1998).

\noindent $\bullet$ In the models predicting either bimodal or wide
DTDs, ``prompt'' type Ia and Core-Collapse (CC) SNe are characterized by
similar delay times and both trace the cosmic star formation history. 
At high redshifts the SNIa ``tardy'' component tends to disappear and 
therefore {\em we predict that the rate ratio CC/Ia steadily increases 
with redshift}, as shown in Figure~2, from a value of about 3 in the
local Universe to about 9 at z$\geq$4. On the contrary, a
``single-population'' model   predicts a much faster evolution of the
CC/Ia ratio, which is expected to become larger than 10 already at
z$\sim$1.5. 

Measuring the SN rates and their CC/Ia ratio at high redshifts  will be
a very interesting task for the  upcoming James Webb Space Telescope and
giant ground-based telescopes  and will permit to verify these
predictions.

%-----------------------------------------------------------------------------

%%%%%%%%%%%%%%%%%%%%%%%%%%%%%%%%%%%%%%%%%%%%%%%%
%% BACKMATTER
%%%%%%%%%%%%%%%%%%%%%%%%%%%%%%%%%%%%%%%%%%%%%%%%

\begin{theacknowledgments}
  NP acknowledges partial support from STScI, through DDRF grant
\#82367, and  from INAF - Observatory of Rome that allowed him to attend
this conference. 
\end{theacknowledgments}

%%%%%%%%%%%%%%%%%%%%%%%%%%%%%%%%%%%%%%%%%%%

\end{document}